\documentstyle[buckow1,12pt]{article}
\newcommand{\be}{\begin{equation}}
\newcommand{\ee}{\end{equation}}
\newcommand{\bea}{\begin{eqnarray}}
\newcommand{\eea}{\end{eqnarray}}
\newcommand{\bi}{\bibitem}
\newcommand{\nn}{\nonumber}
\newcommand{\complex}{{{\rm I} \kern -.59em {\rm C}}}
\begin {document}
\def\titleline{
              $N = 1$ Finite Unified Theories
\newtitleline
                Predictions and Dualities
\footnote{Based on
a talk given at the ``31st International Symposium Ahrenshoop
on the Theory of Elementary Particles''
Buckow, September 2-6, 1997.}
}
\def\authors{
George Zoupanos 
}
\def\addresses{
 Physics Department, National Technical University,\\
 Zografou, GR-15780 Athens, Greece.\\
 and\\
 Institut f.~Physik, Humboldt-Universit\"at,\\
 D10115 Berlin,  Germany
 } 
\def\abstracttext{
$N = 1$, all-loop Finite Unified Theories (FUTs)
 are very interesting not only since they realize an old theoretical
 dream, but also due the remarkable predictive power of particular
 models as well as for providing candidates that might shed light in
 non-perturbative Physics.  Here we discuss (a) the recent developments
 concerning the soft supersymmetry breaking (SSB) sector of these
 theories and the resulting predictions in very interesting
 realistic models, and (b) the results of a recent search for duals of $N
 = 1$, all-loop FUTs.
}
\makefront
\section{Introduction}

Finiteness is an outstanding issue in the various theoretical
endeavors that attempt to achieve a deeper understanding of Nature.
Most theorists, quite independently from the school of thinking they
belong to, tend to believe that the divergencies of ordinary field
theory are not of fundamental nature, but rather they are 
signaling the existence of New Physics at higher scales. Therefore, it
is very natural to believe that the final theory, which hopefully
unifies all interactions, should be completely finite.  Given the
searches of strings, non-commutative geometry and q-groups that aim,
among others, to achieve finiteness, it is worth stressing that {\em
  finiteness does not require necessarily gravity.}  The latter
statement is well established in gauge theories with extended
supersymmetry. For instance, it is well known that all $N = 4$ and
some $N = 2$ supersymmetric gauge theories are free from ultraviolet
(UV) divergencies at all-orders in perturbation theory (PT). Of
particular interest is the existence of \break N = 1 supersymmetric gauge
theories \cite{LPS,acta}, which are
finite at all-orders in PT and the construction in that framework of a
realistic finite $SU(5)$ GUT, which succesfully has predicted, among
others, the top quark mass \cite{kmz1}.

In the following we shall first report on the recent progress that has
been done, and which increases appreciably the predictive power of the
models that can be constructed and provides us with new predictions for
the Higgs particles masses and the s-spectrum, which were not available in
the previous attempts.

The discussion about finite $N = 1$ gauge theories till recently was
limited to perturbative aspects of these theories while
non-perturbative problems, like their bound state spectrum, were left
open. The knowledge of the perturbative properties were sufficient for
the study of GUTs. However, the recent progress that has been done
might permit also studies at strong couplings. The basic idea that
could allow us to address non-perturbative problems in these theories
is the hope that there exist some kind of electric-magnetic,
strong-weak coupling duality, like the one that is believed to be
exhibited in $N = 4$ supersymmetric gauge theories.  Moreover, the low
energy effective action of $N = 2$ supersymmetric gauge theories can
be even solved using duality and holomorphy \cite{SW}.
Finally, $N = 1$ supersymmetric gauge theories, which can describe
realistically the so far observed world, exhibit a weaker version of
duality symmetry \cite{seiberg}, namely theories differing in the UV can
describe the same Physics in the infrared (IR).

In the following we shall also report on a systematic search for duals
of $N = 1$, all-loop finite gauge theories \cite{karch}.
This search covers most of the chiral finite gauge theories including
the realistic $SU(5)$ model mentioned earlier and several vector-like
models. Surpisingly, we found that the duals of the $N = 1$, all-loop
finite gauge theories are usually not finite, with the exception of
an $SO(10)$ theory which emerges from this search as a candidate to
exhibit also S-duality.

\section{Finiteness and Reduction of Couplings in $N=1$ SUSY Gauge Theories}

Consider a chiral, anomaly free,
$N=1$ globally supersymmetric
gauge theory based on a group G with gauge coupling
constant $g$. The
superpotential of the theory is given by
\bea
W&=& \frac{1}{2}\,m^{ij} \,\Phi_{i}\,\Phi_{j}+
\frac{1}{6}\,C^{ijk} \,\Phi_{i}\,\Phi_{j}\,\Phi_{k}~,
\label{supot}
\eea
where $m^{ij}$ and $C^{ijk}$ are gauge invariant tensors and
the matter field $\Phi_{i}$ transforms
according to the irreducible representation  $R_{i}$
of the gauge group $G$. 

 The one-loop $\beta$-function of the gauge
coupling $g$ is given by 
\bea
\beta^{(1)}_{g}&=&\frac{d g}{d t} =
\frac{g^3}{16\pi^2}\,[\,\sum_{i}\,l(R_{i})-3\,C_{2}(G)\,]~,
\label{betag}
\eea
where $l(R_{i})$ is the Dynkin index of $R_{i}$ and $C_{2}(G)$
 is the
quadratic Casimir of the adjoint representation of the
gauge group $G$. The $\beta$-functions of
$C^{ijk}$,
by virtue of the non-renormalization theorem, are related to the
anomalous dimension matrix $\gamma^j_i$ of the matter fields
$\Phi_{i}$ as:
\be
\beta_{C}^{ijk}=\frac{d}{dt}\,C^{ijk}=C^{ijp}\,
~\sum_{n=1}~\frac{1}{(16\pi^2)^n}\,\gamma_{p}^{k(n)} +(k
\leftrightarrow i) +(k\leftrightarrow j)~.
\label{betay}
\ee
At one-loop level $\gamma^j_i$ is 
\be
\gamma_i^{j(1)}=\frac{1}{2}C_{ipq}\,C^{jpq}-2\,g^2\,C_{2}(R_{i})\delta_i^j~,
\label{gamay}
\ee
where $C_{2}(R_{i})$ is the quadratic Casimir of the representation
$R_{i}$, and $C^{ijk}=C_{ijk}^{*}$.

As one can see from Eqs.~(\ref{betag}) and (\ref{gamay}) 
 all the one-loop $\beta$-functions of the theory vanish if
 $\beta_g^{(1)}$ and $\gamma_i^{j(1)}$ vanish, i.e.
\be
\sum _i \ell (R_i) = 3 C_2(G) \,,~~~~~~~~
\frac{1}{2}C_{ipq} C^{jpq} = 2\delta _i^j g^2  C_2(R_i)\,.
\label{2nd}
\ee

A very interesting result is that the conditions (\ref{2nd}) are
necessary and sufficient for finiteness at
the two-loop level.
 
A natural question to ask is what happens at higher loop orders.  The
finiteness conditions (\ref{2nd}) impose relations between gauge and
Yukawa couplings.  We would like to guarantee that such relations
leading to a reduction of the couplings hold at any renormalization
point.  The necessary, but also sufficient, condition
for this to happen is to require that such relations are solutions to
the reduction equations (REs) 
\be \beta_g {d C^{ijk}\over dg} = \beta^{ijk}
\label{redeq2}
\ee
and hold at all-orders.  Remarkably the existence of
all-order power series solutions to (\ref{redeq2}) can be decided at
the one-loop level \cite{zim1}.

A very interesting theorem \cite{LPS} guarantees the
vanishing of the $\beta$-functions to all-orders in perturbation theory,
if we demand reduction of couplings, and that all one-loop anomalous
dimensions of the matter fields in the completely and uniquely reduced
theory vanish identically.

\section{All-Loop Finite Unified Theories}

A predictive Finite Unified $SU(5)$ model which is finite to all-orders,
in addition to the requirements mentioned already, should also
have the following properties:

\begin{enumerate}

\item 
One-loop anomalous dimensions are diagonal,
i.e.,  $\gamma_{i}^{(1)\,j} \propto \delta^{j}_{i} $.

\item
Three fermion generations, $\overline{\bf 5}_{i}~~
(i=1,2,3)$, obviously should not couple to ${\bf 24}$.
This can be achieved for instance by imposing $B-L$ 
conservation.

\item
The two Higgs doublets of the MSSM should mostly be made out of a 
pair of Higgs quintet and anti-quintet, which couple to the third
generation.
\end{enumerate}
In the following we discuss two versions of the all-order finite
model.

\vspace{0.2cm}
\noindent
${\bf A}$:  The model of ref. \cite{kmz1}.
\newline
${\bf B}$:  A slight variation of  the 
 model ${\bf A}$.

The  superpotential which describe the two models 
takes the form \cite{kmz1,kkmz-npb}
\bea
W &=& \sum_{i=1}^{3}\,[~\frac{1}{2}g_{i}^{u}
\,{\bf 10}_i{\bf 10}_i H_{i}+
g_{i}^{d}\,{\bf 10}_i \overline{\bf 5}_{i}\,
\overline{H}_{i}~] +
g_{23}^{u}\,{\bf 10}_2{\bf 10}_3 H_{4} \\
 & &+g_{23}^{d}\,{\bf 10}_2 \overline{\bf 5}_{3}\,
\overline{H}_{4}+
g_{32}^{d}\,{\bf 10}_3 \overline{\bf 5}_{2}\,
\overline{H}_{4}+
\sum_{a=1}^{4}g_{a}^{f}\,H_{a}\, 
{\bf 24}\,\overline{H}_{a}+
\frac{g^{\lambda}}{3}\,({\bf 24})^3~,\nn
\label{super}
\eea
where 
$H_{a}$ and $\overline{H}_{a}~~(a=1,\dots,4)$
stand for the Higgs quintets and anti-quintets.

The non-degenerate and isolated solutions to $\gamma^{(1)}_{i}=0$ for
the models $\{ {\bf A}~,~{\bf B} \}$ are: 
\bea (g_{1}^{u})^2
&=&\{\frac{8}{5},\frac{8}{5} \}g^2~, ~(g_{1}^{d})^2
=\{\frac{6}{5},\frac{6}{5}\}g^2~,~
(g_{2}^{u})^2=(g_{3}^{u})^2=\{\frac{8}{5},\frac{4}{5}\}g^2~,\label{SOL5}\\
(g_{2}^{d})^2 &=&(g_{3}^{d})^2=\{\frac{6}{5},\frac{3}{5}\}g^2~,~
(g_{23}^{u})^2 =\{0,\frac{4}{5}\}g^2~,~
(g_{23}^{d})^2=(g_{32}^{d})^2=\{0,\frac{3}{5}\}g^2~,
\nn\\
(g^{\lambda})^2 &=&\frac{15}{7}g^2~,~ (g_{2}^{f})^2
=(g_{3}^{f})^2=\{0,\frac{1}{2}\}g^2~,~ (g_{1}^{f})^2=0~,~
(g_{4}^{f})^2=\{1,0\}g^2~.\nn 
\eea 
According to the theorem of
ref.~\cite{LPS} these models are finite to all-orders.  After the
reduction of couplings the symmetry of $W$ is enhanced
\cite{kkmz-npb}.

The main difference of the models
${\bf A}$ and ${\bf B}$ is
that three pairs of Higgs quintets and anti-quintets couple to 
the ${\bf 24}$ for ${\bf B}$ so that it is not necessary 
to mix
them with $H_{4}$ and $\overline{H}_{4}$ in order to
achieve the triplet-doublet splitting after the symmetry breaking 
of $SU(5)$.

\section{Supersymmetry Breaking and Predictions of Low Energy Parameters}

 The above models are completed as realistic theories by adding SSB terms
as follows: 
\bea
-{\cal L}_{\rm SB} &=&
\frac{1}{6} \,h^{ijk}\,\phi_i \phi_j \phi_k
+
\frac{1}{2} \,b^{ij}\,\phi_i \phi_j
+
\frac{1}{2} \,(m^2)^{j}_{i}\,\phi^{*\,i} \phi_j+
\frac{1}{2} \,M\,\lambda \lambda+\mbox{h.c.}~,
\eea
where the $\phi_i$ are the
scalar parts of the chiral superfields $\Phi_i$ , $\lambda$ are the gauginos
and $M$ their unified mass.

 Concerning the supersymmetry breaking sector of the theory it has been
recently shown that the requirement of two-loop finiteness of SSB in a
finite gauge theory leads to

\begin{enumerate}
\item the relation $h^{ijk} = - M Y^{ijk}$ and,
\item the soft scalar-mass sum rule
  $( m_i^2 + m_j^2 + m_k^2 )/M M^{\dagger} = 1 + g^2/16\pi^2 \Delta^{(1)}$
for i,j,k with $C^{ijk} \ne 0$, where $\Delta^{(1)}$ is the two-loop
correction, which vanishes for the universal choice \cite{jack3}, but
  also in the considered models without universal masses.   
\end{enumerate}

Since the gauge symmetry is spontaneously broken
below $M_{\rm GUT}$, the finiteness conditions 
do not restrict the renormalization property at low energies, and
all it remains are boundary conditions on the
gauge and Yukawa couplings (\ref{SOL5}),
 the $h=-MC$ relation  and
the soft scalar-mass sum rule  at $M_{\rm GUT}$.
We have examined the evolution of these parameters according
to their renormalization group equations at two-loop 
for dimensionless parameters and 
 at one-loop 
for dimensional ones with
these boundary conditions.
Below $M_{\rm GUT}$ their evolution is assumed to be
governed by the MSSM. It is further assumed a unique 
supersymmetry breaking scale
$M_{s}$ so that
below $M_{s}$ the SM is the correct effective theory.

The predictions for the top quark mass $M_t$ are $\sim 183$ and $\sim
174$ GeV in models $\bf A$ and $\bf B$ respectively. Comparing these
predictions with the most recent experimental value $ M_t = (175.6 \pm
5.5)$ GeV, and recalling that the theoretical values for $M_t$ may
suffer from a correction of less than $\sim 4 \%$ \cite{acta}, we see
that they are consistent with the experimental data.

 Using the sum rule we can now determine the
spectrum of realistic models in terms of just a few parameters. In
addition to the successful prediction of the top quark mass the
characteristic features of the spectrum are that 1) the lightest Higgs
mass is predicted $\sim 120$ GeV and 2) the s-spectrum starts above
200 GeV. Therefore, the next important test of these Finite Unified
Theories will be given with the measurement of the Higgs mass, for
which the models show an appreciable stability in their prediction.

\section{Dualities of Finite Gauge Theories}

In ref.~\cite{karch} we have done an extensive search for dual
gauge theories of all-loop finite, $N = 1$ supersymmetric gauge
theories.  It is shown how to find explicitly the dual gauge theories
of almost all chiral, $N = 1$, all-loop finite gauge theories, while
several models have been discussed in detail, including the realistic
finite $SU(5)$ unified theory of ref. \cite{kmz1}. As we
have seen one- and two-loop finiteness of a gauge theory is guaranteed by,
first, choosing the particle content such that the one-loop gauge
$\beta$-function vanishes, and subsequently, by adding a superpotential
such that all one-loop matter field anomalous dimensions are zero.
Furthermore, the all-loop finiteness requires that the relations among
gauge and Yukawa couplings, obtained by demanding the vanishing of the
one-loop anomalous dimensions, should be unique solutions of the
reduction equations, which in turn guarantees that they can be
uniquely determined to all-orders in PT. Therefore, using established
methods for searching for duals \`a la Seiberg, we have examined almost
all known $N = 1$ supersymmeric gauge theories (with the exception of
$E_6$ models and $SO(10)$ containing anti-spinors) and vanishing one-loop
$\beta$-functions. These theories have, first, been promoted to all-loop
finite ones, by adding appropriate superpotential and then by meeting
the requirements of all-loop finiteness. In addition, certain
vector-like, all-loop finite, $N = 1$ gauge theories and their duals
have been discussed in the standard field theory framework but also by
using the derivation of gauge theories from branes. However, the brane
picture still encounters several difficulties in the corresponding
hunting for finite gauge theories.

The result of our search is that the duals \`a la Seiberg of all-loop
finite gauge theories are asymptotically free. In certain cases
looking to the IR limit of the theories, where both theories of the
dual pair are describing the same physics, we found that the
spontaneously broken theories seem superficially asymptotically
non-free (as the $U(1)$ of the Standard Model embedded in a GUT).

There is only one model which is one-loop finite in the IR. The model
is based on the gauge group $SO(10)$ and has matter content consisting
of 8 vector and 8 spinor superfields. The dual of this theory is based
on the gauge group $SU(17)\times SP(14)$. The first factor is
asymptotically free, while the second is one-loop finite. However, in the
IR, after spontaneous symmetry breaking of the $SU(17)$ down to $SU(9)$,
both gauge factors are one-loop finite. We found evidence that the IR
dual is all-loop finite and therefore the dual pair is a candidate to
exhibit $N = 1$ S-duality.
 
The dual of the realistic finite unified theory \cite{kmz1} based on
$SU(5)$ has been determined and discussed in some detail. However, the
resulting dual theory is rather complicated and it does not give, so
far, any hint for any useful use of it. On the other hand, we should
note that the dual was constructed by using the deconfinement method
\cite{berkooz}, which does not lead to unique results.  Therefore, we
should not exclude the possibility that more interesting duals can be
constructed.

\section{Conclusions}

The search for realistic Finite Unified theories started a few years ago
\cite{kmz1,acta} with the successful prediction of the top quark mass,
and it has now been complemented with a new important ingredient
concerning the finiteness of the SSB sector of the theory. 
Specifically, a sum rule for the soft scalar masses has been obtained
which quarantees the finiteness of the SSB parameters up to two-loops
\cite{kkmz-npb}, avoiding at the same time serious phenomenological
problems related to the previously known ``universal'' solution.  It is
found that this sum rule coincides with that of a certain class of
string models in which the massive string modes are organized into
$N=4$ supermultiplets.

Motivated from the recent developments on dualities of gauge theories
with extended supersymmetry we have been searching for candidates
with S-duality among the $N = 1$ supersymmetric gauge theories, which
have richer dynamics and are much more promissing in describing the
real world, as the one discussed above.  The strategy was to look for
duals \`a la Seiberg of all-loop, $N = 1$ FUTs which are FUTs too and
therefore exhibiting S-duality. From our search, one chiral, $N = 1$, 
all-loop FUT has been singled out giving promises of having S-duality
as the $N = 4$ gauge theories.

{\it Added notes}: Since the submission of the present contribution there exist
some interesting related developments, which we would like to
mention. The first concerns the soft scalar-mass rule presented in
sect. 4, which has been extended recently to all-orders \cite{kkz}. In the
case of finite unified theories, the sum rule ensures the all-loop
finiteness in the soft supersymmetry breaking sector. In addition the
exact $\beta$-function for the soft scalar masses in the NSVZ scheme
\cite{NSVZ} for softly-broken supersymmetric QCD has been obtained
\cite{kkz}.
The second concerns the construction of a $N=1$, one-loop finite $SU(4)^3$
gauge theory with 3 chiral generations in three dual ways from strings
\cite{iba}. In addition, motivated by the connection among gauge theories
realized on branes and their supergravity realizations at large $N$
\cite{mald}, a generalization of the above model has been constructed,
by studying D3 branes at orbifolds, based on $U(N)^3$ gauge group with
three chiral generations \cite{ks}. The latter models appear
together with a superpotential and it can be easily shown that are 
all-loop finite using the methods that have been reviewed here. For both
models naturally it is conjectured that they have improved properties such as
S-duality since they are directly connected to or hoped to have inherited
properties from $N=4$ gauge theories. It is very interesting to see what
can really be proven.

\section{Acknowledgements}

It is a pleasure to thank the Organizing Committee for
the warm hospitality.  Work partially supported  by the projects
FMBI-CT96-1212 and ERBFMRXCT960090, and  PENED95/1170;1981.

\end{document}